\documentclass[structabstract]{aa} 
\usepackage{natbib}
\usepackage{graphicx}
\usepackage{mathtools}

\begin{document}

\title{ELECTRON ATTACHMENT RATES FOR PAH ANIONS IN THE ISM AND DARK MOLECULAR CLOUDS:
DEPENDENCE ON THEIR CHEMICAL PROPERTIES} 

\titlerunning{Electron Attachment Rates For PAH Anions}

\author{F.Carelli\inst{1} \and T.Grassi\inst{1} \and F.A. Gianturco\inst{1}\thanks{Corresponding author: e-mail: fa.gianturco@caspur.it Fax +39-06-49913305}}

\institute{Department of Chemistry and CISM, The University of Rome `Sapienza', P.le Aldo Moro 5, 00185 Rome, Italy.}

\date{Received *****; accepted *****}

\newcommand{\dd}{\mathrm{d}}
\newcommand{\kbol}{k_\mathrm{B}}


\abstract
{The attachment of free electrons to polycondensed aromatic ring molecules (PAHs) is studied for the variety of these molecules with different numbers of condensed rings and over a broad range of electron temperatures, using a multichannel quantum scattering approach. The calculations of the relevant cross sections are used in turn to model the corresponding attachment rates for each of the systems under study, and these rates are parametrized as a function of temperature using a commonly employed expression for two-body processes in the interstellar medium (ISM).}
{The scope of this work is to use first principles to establish the influence of chemical properties on the efficiency of the electron-attachment process
for PAHs.}
{Quantum multichannel scattering methods are employed to generate the relevant cross sections, hence the attachment rates, using integral elastic cross sections computed over a broad range of relevant energies, from threshold up to 1000 K and linking the attachment to low-energy resonant collisions.}
{The rates obtained for the present molecules are found to markedly vary within the test ensemble of the present work and to be lower than the earlier values used for the entire class of PAHs anions, when modelling their evolutions in ISM environments.
The effects of such differences on the evolutions of chemical networks that include both PAH and PAH- species are analysed in some detail and related to previous calculations.}{}

\maketitle

\section{Introduction}
\label{intro} 

\noindent The widespread presence in the interstellar medium (ISM) of polycyclic aromatic hydrocarbons (PAHs), i.e. organic molecules made up of several aromatic rings fused together, as well as the existence in the same environments of their dehydrogenated, ionized, and/or protonated counterparts, has been largely inferred from the extensive presence of unidentified infrared emission bands that have been observed in the range of 3 to 14 micrometres \citep{Hendecourt1997, Rhee2007,Parker2012,Ricks2009}. Current astrochemical models for the PAH formation have largely been derived from the combustion chemistry community that suggest how these formation processes should occur in warm and dense circumstellar envelopes of carbon-rich stars through sequential reactions of smaller radicals with acetylene at temperatures around 1000 K (\citet{Cherchneff1992} and erratum).
Even more recently, their formation at temperatures down to 10 K has been surmised after an interesting investigation that combines cross molecular beam studies and quantum chemistry calculations \citep{Parker2012}.

Another low-energy path to possible formations of PAHs has come from investigating the role of the free electrons that are present in the circumstellar and interstellar media and in protoplanetary atmospheres \citep{Herbst2008} in order to assess from calculations and possible experimental data the likely attachment efficiency of these free projectiles to the PAHs and the ensuing formation of very reactive anionic species, which can in turn react with radicals present in those environments \citep{DeMarais2012}.
That even temporary anions of smaller members of this family of aromatic molecules could be involved in subsequent reactions with cationic radicals has also been suggested in our earlier computational studies of benzyne anions \citep{Carelli2010, Carelli2011}. Recent experimental work has shown that indeed  anionic PAHs like phenide (C$_6$H$_5^-$), naphthalenide (C$_{10}$H$_7^-$), and anthracenide (C$_{14}$H$_9^-$) can efficiently react with H, H$_2$, and D$_2$ as observed in flowing afterglow-selected experiments \citep{DeMarais2012,Yang2011}.
It therefore follows from the above that to establish the possible effects of the presence of anionic PAHs on the chemistry of dark circumstellar regions  and to model both the possible efficiency of formations and the consequences of such additional partners on the evolutionary history of that chemistry becomes directly interesting for improving our understanding of the PAH chemistry in the ISM in general. In particular, the present work wishes to analyse whether the chemical properties of the individual component partners in the large series of postulated PAHs play any significant role in the evolutionary studies of the chemical networks and if it is still a realistic choice to view their efficiency as given by a single rate value with a unique temperature dependence.

In the following we therefore intend to approach the problem of identifying, if at all possible, differences in behaviour for a subset of PAHs chosen as initial examples for that entire class of molecules, which are related to their differences in chemical properties. 
In the next section we therefore summarize first the methods employed in earlier models for obtaining the attachment rates for the PAHs and then present our quantum scattering calculations in some detail to obtain integral cross sections from threshold up to several eV of energy and further link such observables with the model we use to yield the relevant attachment rates. Section 3 reports our results for a small group of PAH molecules and discusses their differences in relative efficiencies, while section 4 describes the final effects of such differences on the evolutionary history of the chemical networks currently employed for simulations. Our conclusions are given in section 5.

\section{Computing cross sections and attachment rates}
\label{sect_comp_cross}

\subsection{Earlier evaluations of the rates}
The importance of evaluating the efficiency of electron attachment to PAHs under ISM conditions has been recognized very early on 
(e.g. see \citealt{Herbst1981,Petrie1997}) because of the enhanced role in the network kinetics that the presence of anionic reagents can induce during evolutionary modelling of proto-planetary and stellar objects \citep{Bass1979}. The earlier estimates \citep{Herbst1981}, however, did not carry out explicit quantum dynamical analysis of the interaction and attachment mechanism of the impinging electron, but rather argued the possibility of applying phase-space theory to generate vibrational density of states for final species, the stable anion M$^-$, at the scaled internal energy corresponding to the electron affinity value (EA) for the initial species, the neutral partner M. 
The above treatment involved a careful analysis based
on the Rice-Ramsperger-Kassel-Marcus (RRKM) treatment \citep{Marcus1956} of the electron-attachment kinetics whereby
the vibrational structures of the ``reagent'' (the initial neutral molecule) and of the ``products''
(the anionic stable molecules) are estimated from structural calculations and are employed to
produce the final rates of formation. The conclusion was that radiative attachment of electrons to neutrals could occur efficiently at the low-temperature of the ISM for polyatomic neutrals containing more than three to four atoms and having EA values higer than 2-3 eV \citep{Herbst2008}. The above theory was discussed further by \citet{Herbst1985} and \citet{Petrie1997}, where the conjecture was added that the attachment process is efficiently initiated primarily by s-wave impinging electrons.

One should stress at this point that those previous studies of the electron attachment rates to linear polyynes (e.g. see
\citealt{Herbst2008} and the reference quoted therein) also follow phase-space theory reasoning along with the dominance of s-wave electron attachment at 
thresholds, although no dynamical coupling between nuclear vibration and impinging electrons has ever been explicitly included in that modelling. The approach of the RRKM and conservation of angular momentum \citep{Pechukas1965}, therefore, involves (i) the conjecture that electron in s-wave are captured at vanishing energies, and (ii) that a phase-space analysis of the neutral molecule's vibrational levels, \emph{vis-\`a-vis} the vibrational density of states for the final, stable anion is employed to estimate attachment rates. Neither assumption, however,  originates in specific and explicit dynamical
calculations, and therefore no detailed computations of electron-molecule scattering cross sections
have been employed thus far to generate the possible attachment rates of the scattered electrons to the relevant target molecules. In the
following, we instead present an evaluation of such rates starting from \emph{ab initio} scattering calculations of the cross sections. 

\subsection{The quantum dynamics for cross sections}
\label{model1}
The method we have employed to generate the final rates is based on quantum calculations from ab initio methods of the necessary integral cross sections.
This has been described in greater detail in our earlier work - e.g. see \citet{Lucchese1996} - so we only give a brief
outline here of its most relevant features.
In our method both the target one-electron bound orbitals and the impinging electron wavefunction are expanded as linear combination of symmetry-adapted spherical harmonics

\begin{equation}
\label{sce}
\phi _{i}^{p \mu} (r,\theta,\phi) = \frac{1}{r} \sum _{lh} u_{lh}^{i,p \mu} (r) \chi _{lh}^{p \mu}(\theta,\phi)\,
\end{equation}
where the $\chi _{lh}^{p \mu}(\theta,\phi)$ indicate the linear combinations of spherical harmonics $Y_{l}^{\mu}$, and the $|p\mu \rangle$ labels record the specific irreducible representation (IR) indices.
The assumption that the target molecule is adequately described by its ground-state Slater determinant (static exchange approximation) leads to the following simplified form for the scattering equations:

\begin{equation}
\label{ide}
\left[\frac{1}{2}\nabla ^{2} +(E-\epsilon)\right]F({\bf r}) = \int dr'V({\bf r,r'})F({\bf r'})\,,
\end{equation}

\noindent that can be solved (once the model interaction potential for our system is assembled) by using the Schwinger variational method as described in our earlier publications, e.g. see \citet{Lucchese1996}.
The static exchange approximation does not include the response of the target to the impinging electron, i.e. the correlation and polarization effects acting at short and at large electron-target distances, respectively.
That additional part of the overall interaction has been modelled in our study by writing

\begin{equation}
\label{vcp}
  V^{cp}({\bf r}) = \left\{
  \begin{array}{l l}
    V^{corr}({\bf r}) & \quad \text{for} \,\bf r\leq\bf r_\mathrm{match}\\
    V^{pol}({\bf r}) & \quad  \text{for} \,\bf r >  \bf r_\mathrm{match}\,.\\
  \end{array} \right.
 \end{equation}

\noindent In the present work we have used the Lee-Yang-Parr form for $V^{corr}({\bf r})$ \citep{Lucchese1996} and further employed the $V^{pol}({\bf r})$ as a function of the polarizability tensor to describe successfully, via the $V^{cp}$ model of Eq.(\ref{vcp}), this additional part of the overall interaction potential.
We further replace the exact non-local exchange interaction of Eq.(\ref{ide}) with a local energy-dependent potential.

\begin{equation}
\label{hara}
V_{FEGE} = \frac{2}{\pi}k_{F}({\bf {r}}) \left( \frac{1}{2} + \frac{1-\eta^{2}}{4\eta}\ln\left|\frac{1+\eta}{1-\eta}\right|\right)
\end{equation}

\noindent where $k_{F}$ is the Fermi momentum and $\eta$ the neutral molecule ionization potential.
We have already found  \citep{Lucchese1996} that Eq.(\ref{hara}) can yield a reasonably realistic description of such forces in a wide variety of molecular systems.
We can rewrite the scattering equations in homogeneous form to obtain the static-model-exchange-correlation-polarization (SMECP) approximation for the scattering event

\begin{equation}
\label{smecp}
\left[-\frac{1}{2}\nabla^{2}-\frac{1}{2}k^{2}+\overbrace{\hat V^{st}+\hat V^{cp} +\hat V^{FEGE}}^{V(\textbf{r})}\right]F({\bf r}) = 0\,.
\end{equation}

\noindent Here, $V({\bf r})$ is the sum of the three local potentials $V({\bf r}) = V^{st}({\bf r}) + V^{cp}({\bf r}) + V^{FEGE}({\bf r})$, of which $V^{st}({\bf r})$ describes the static interaction of the scattered electron with the target's nuclei and electrons, $V^{cp}({\bf r})$ contains the correlation-polarization contributions of Eq.(\ref{vcp}), and $V^{FEGE}({\bf r})$ is the model exchange potential of Eq.(\ref{hara}).
After integrating over the angular variables, Eq.(\ref{smecp}) takes the form

\begin{equation}
\label{idebis}
\left[\frac{d^{2}}{dr^{2}}-\frac{l(l+1)}{r^{2}}+k^{2}\right]F^{p\mu}_{lh}(r)=2\sum_{l'h'}V_{lh,l'h'}^{p\mu}(r)F_{l'h'}^{p\mu}(r)\,,
\end{equation}

\noindent where the potential coupling elements are given by

\begin{eqnarray}
V_{lh,l'h'}^{p\mu}(r) &=& \langle\chi_{lh}^{p\mu}(\hat r)|V({\bf r})|\chi_{l'h'}^{p\mu}(\hat r)\rangle = \nonumber\\ 
	&=& \int d\hat r \chi_{lh}^{p\mu}(\hat r)V({\bf r})\chi_{l'h'}^{p\mu}(\hat r)\,.\label{potels}
\end{eqnarray}

\noindent The S-matrix elements, related to the K-matrix, yield the set of partialwave phaseshifts

\begin{equation}
\label{smatrix}
S_{l} = S_{l}(k) = e^{2i\delta_{l}(k)}
\end{equation}

\noindent for all the contributing angular momenta.
For a scattering process characterized by many broad and overlapping resonances, the usual Breit-Wigner formula is no longer efficient.
However it is possible to extract the positions and the widths of closely spaced resonant states by using the Q-matrix formalism

\begin{equation}
Q(E) = i\hbar S\frac{dS^{\dagger}}{dE} = -i\hbar \frac{dS}{dE}S^{\dagger} = Q^{\dagger} (E)\,.
\end{equation}

\noindent Eigenvalues of the $Q$ matrix are equal to the delay time of the outgoing wavepacket, as defined in \citet{Smith1960}. For a multichannel scattering process the trace of the Q matrix is further related to the eigenphase sum $\delta_{sum}$,

\begin{equation}
2\hbar \frac{d\delta_{sum}}{dE} = Tr Q(E)\,
\end{equation}

\noindent where the eigenphase sum is defined via the elements of Eq.(\ref{smatrix}), with $k^{2} = 2E$

\begin{equation}
\delta_{sum} = \sum_{l=1}^{\infty} \delta_{l}(E)\,.
\end{equation}

\noindent All the necessary partialwave contributions obtained at each collision energy are subsequently employed to yield the final integral cross sections (ICS) for each specific molecular target.

In conclusion, our \emph{ab initio} scattering calculations allow us to obtain locations and lifetime of transient negative ion (TNI) complexes,
to analyse the presence of virtual states at zero-energy enhancing TNI formation, and to locate the role of dipole-bound states for dipolar molecules
by analysing the scattering wavefunctions at vanishinging low energies. All such features will play a role, as described below, in our final estimates
of the corresponding attachment rates. 

\subsection{Generating attachment rates from cross-sections}
The next step in the calculations involves the thermal average over the internal states of the species in the 
two-body (2B) reactions that will be analysed below. By assuming that the free electrons have a kinetic temperature
$T_e$ and that the molecules interacting with them have an internal distribution over states that can be associated with a 
temperature\footnote{Not to be confused with the kinetic molecular temperature.} $T_\mathrm{M}$, the existence of local thermal equilibrium (LTE) implies, given the large differences between the electron's mass and those
of the molecules, that $T_\mathrm{M}\neq T_e=T$ (i.e. the molecules are considered as still during collisions and 
being in their ground rotovibrational states), thus the rate of formation of molecular anions is given by
\begin{equation}
k_\mathrm{TNI}(T)=\frac{1}{Q_\mathrm{v}Q_\mathrm{r}Q_\mathrm{el}}\sum_n e^{-\epsilon/\kbol T}\int v\sigma_n(v)f(v,T)\dd v\,,
\end{equation}
where $\sigma_n(v)$, a function of the relative velocity, is the integral cross-section (ICS) for electron-molecule
collision labelled by $n$, a global index that specifies the electronic, vibrational and rotational initial
molecular states, and the $Q$'s are their corresponding partition functions. In the present study we limit the target
description to be sufficient for generating ICS which are summed over all final rotational states accessible at the considered energy (and averaged over all initial states as taken to be $|j_0\rangle=0$). The molecule is also
kept in its vibrational and electronic ground states given the likely conditions in the ISM environment. If the
$f(v,T)$ is taken to be the standard Maxwell-Boltzmann electron's velocity distribution, then we can write
\begin{equation}\label{2Bkform}
	k_\mathrm{TNI}(T) = \left(\frac{8\kbol T_e}{\pi\mu_e}\right)^{1/2}\frac{1}{(\kbol T_e)^2}
		\int E\sigma_0(E)e^{-E/\kbol T}\dd E\,,
\end{equation}
which describes the rate of metastable anionic formation via 2B processes and with molecules in their ground electronic, 
rotovibrational states. The total cross section $\sigma_0$ of Eq.(\ref{2Bkform}) is obviously made up of different processes that have to be taken into considerations when discussing the final formation of a stable molecular anion.

As a result, the molecular processes that are possible via binary collisions are more specifically described as those initially leading to anionic metastable species formation before energy redistribution, i.e. leading to TNI formation in the continuum of electron energies
\begin{equation}\label{TNIproc}
\mathrm{M_{eq}}+\mathrm e^-\xrightarrow{k_\mathrm{TNI}}[\mathrm{M_{eq}}^-]^*\,.
\end{equation}
Each of such TNI corresponds, at each specific energy window, to a resonant process of attachment before molecular relaxation into a bound anion M$^-$. The dynamical channels of the collisions of incoming electron with the molecule also involve direct scattering occurring over the whole
energy range. 
Both processes are explicitly considered in the present calculations, which therefore locate all the shape-resonant states (TNI) formed over the range of collision energy considered, establish their effects on changing the size of the integral cross sections, and also provide a realistic calculation of the background cross sections over the whole energy range. Given the high polarizabilities exhibited by the present molecules, and especially when permanent dipole moments are present, the calculations go down to threshold energies and therefore also provide a realistic description of virtual states and s-wave attachment effects which are very important for describing regions of efficient couplings between the incoming electron and the nuclear network of the target molecule \citep{Carelli2012b}. 

As a consequence of this coupling between the attached electron and the molecular vibrations, we can further note that the most relevant relaxation processes following Eq.(\ref{TNIproc}) are given by (i) radiative  stabilization (RS) dynamics in the chiefly collisionless ISM environment
\begin{equation}\label{RSproc}
[\mathrm{M_{eq}}^-]^*\xrightarrow{k_\mathrm{RS}}\mathrm{M_{eq}}^-+h\nu\,,
\end{equation}
in competition with (ii) the autodetachment (AD) process before anionic stabilization
\begin{equation}\label{ADproc}
[\mathrm{M_{eq}}^-]^* \xrightarrow{k_\mathrm{AD}}\mathrm{M}+\mathrm e^-\,.
\end{equation}

Current assumptions consider $k_\mathrm{AD}\gg k_\mathrm{RS}$ \citep{Herbst1981}, which allows us to simplify the
expression for the overall formation rate $k_f(T)$ of Eq.(\ref{2Bkform})
\begin{equation}
k_f(T) = \left(\frac{k_\mathrm{RS}}{k_\mathrm{AD}+k_\mathrm{RS}}\right)k_\mathrm{TNI}
\end{equation}
into the form
\begin{equation}
k_f(T) = \left(\frac{k_\mathrm{RS}}{k_\mathrm{AD}}\right)k_\mathrm{TNI}\,.
\end{equation}
The above relation then tells us that the $k_\mathrm{TNI}$ rates can provide an \textbf{upper} bound to the true $k_f(T)$
insofar as the radiative stabilization remains less efficient than the autodetachment process, so that we can safely
assume that
\begin{equation}\label{kf_le_kTNI}
k_f(T)\le k_\mathrm{TNI}(T)\,.
\end{equation}
What we are essentially saying with Eq.(\ref{kf_le_kTNI}) is that the present ensemble of mechanisms that
can lead to a stable molecular anion as a final product (described by the rate $k_f$) require the presence of (i) resonant attachment at various low energies but at specific energy windows, and (ii) virtual state formations at the energy thresholds (zero-energy attachment). Both the above processes are included in our scattering calculations that lead to the $k_\mathrm{TNI}$ rates. The next step requires the molecular TNI relaxation to a bound anion and would need to explicitly include the nuclear dynamics during the scattering process, i.e. explicit dissociative electron attachment (DEA) and/or vibrational Feshbach resonances (VFR) treatments.
Since we carry out calculations at a fixed nuclear geometry, then the $\sigma_0$ computed for Eq.(\ref{2Bkform}) corresponds to all cross sections leading to attachment and therefore to taking $k_f\approx k_\mathrm{TNI}$.

By using the computed rates of Eq.(\ref{2Bkform}) the presently calculated $k_\mathrm{TNI}$, thus provides an
evaluation of the electron-attachment probabilities by considering that all the occurring collisions
will lead to efficient, vibrationally-driven decays into bound anionic states. They therefore
should give upper limits to the anionic formation rates of PAH- species in interstellar environments
and do so by using \emph{ab initio} calculations of the relevant scattering cross sections.

\section{Our computed attachment rates}
In the present, exploratory analysis we have considered a series of molecules that can help us to establish the extent of chemical diversity existing among possible PAH candidates. In particular, we chose two single-ring species,
benzyne (C$_6$H$_4$) and phenil (C$_6$H$_5$) because of their having a permanent dipole moment. Thus, although
they are often considered precursor molecules to the PAHs \citep{Widicus2007}, we analyse here their rates chiefly
to test the relevance of their being polar molecular targets.

As examples of polyconjugated systems, we report the attachment rates computed for anthracene (three rings: 
C$_{14}$H$_{10}$), perylene (five rings: C$_{20}$H$_{12}$), and coronene (seven rings: C$_{14}$H$_{12}$).
The full details of the cross section calculations have already been published for the coronene \citep{Carelli2012b},
and are in the process of being piblished for anthracene \citep{Sanz2012} and perylene \citep{Carelli2012c}. For the single-ring
molecules, the benzyne data have already been published \citep{Carelli2010,Carelli2011} while the phenil cross-sections are being
prepared for publication \citep{Carelli2012a}. All the calculations have followed the computational method described briefly in the previous section (\ref{sect_comp_cross}).

For the 2B reactions, the expression for the rate coefficient is usually - e.g. see \citet{Woodall2007} -
written down in the following form to model unimolecular processes, where electron attachment processes are also examples,
\begin{equation}\label{fitrate}
	k_f=\alpha\left(\frac{T}{300\,\mathrm{K}}\right)^\beta \exp(-\gamma/T)\,\mathrm{cm}^3\mathrm{s}^{-1}\,,
\end{equation}
so that the necessary information for a specific reaction is stored in the ($\alpha$,$\,\beta$,$\,\gamma$) set of
parameters identifying the temperature dependence of its rate coefficient for that particular PAH.

\begin{table}
	\caption{Fit coefficients for the moelcules presented in this paper, where 
		$\alpha$ is in cm$^3$ s$^{-1}$, $\beta$ is dimensionless, and $\gamma$ in K. Note that $a(b)=a\times10^b$. 
		See text and Eq.(\ref{fitrate}) for further details.}
	\centering
	\begin{tabular}{l|rrr}
		\hline
		 & $\alpha$ & $\beta$ & $\gamma$\\
		\hline
		C$_6$H$_4$ & 10.02(-9) & -0.33 & 6.69\\
		C$_6$H$_5$ & 1.44(-9) & -0.12 & -14.77\\
		C$_{14}$H$_{10}$ & 1.31(-9) & 0.21 & -1.12\\
		C$_{20}$H$_{12}$ & 3.06(-9) & -0.01 & 2.77\\
		C$_{24}$H$_{12}$ & 2.74(-9) & 0.11 & -1.12\\
		\hline
	\end{tabular}
\label{tab_abc}
\end{table}

\begin{figure}
\begin{center}
\includegraphics[width=.35\textwidth, angle= 0]{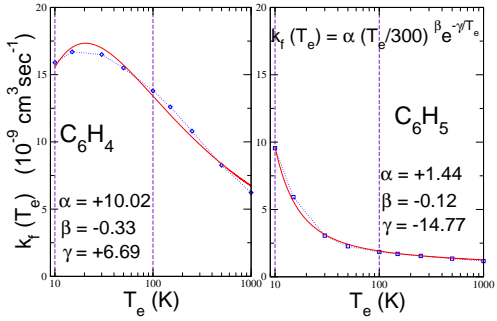}
\caption{Computed electron capture rates for phenil and benzyne polar aromatic molecules. The corresponding fitting parameters for 
	Eq.(\ref{fitrate}) are also given in the figure. Here, $\alpha$ is in units of $10^{-9}$ cm$^3/$s.\label{fig1}}
\end{center}
\end{figure}

The calculations of Fig.\ref{fig1} report, as specific examples, the electron-attachment rates for phenil and benzyne,
over a significant range of temperatures: the parameters from the fitting suggested by Eq.(\ref{fitrate})
are also given for each system (see Table \ref{tab_abc}).

The calculated dipole moment of C$_6$H$_4$ is $1.38$ D \citep{Kraka1993}, while the one of C$_6$H$_5$ is $0.9$ D
\citep{McMahon2003}; all the remaining polycondensed examples of the present work have no dipole moment.
What one clearly sees from the data in Fig.\ref{fig1} is the strong increase in both rates at the threshold 
temperatures. Both polar targets indicate more efficient attachment for very slow electrons, since both systems have been 
shown to form virtual states at zero energy and also dipole-bound states for the extra electron attached at very low
collision energy \citep{Carelli2010,Carelli2011,Carelli2012a}. Furthermore, we see that the rates for such PAH precursors vary markedly among
themselves and follow the size of the permanent dipole. The latter is larger for benzyne and thus the corresponding
rates for this molecule are a factor of two more than those for phenil at the energy thresholds.
Both rates turn out to be of the order of $10^{-8}$ cm$^3$ s$^{-1}$ in the range of temperatures of
relevance for ISM conditions (i.e. up to about 100 K), although we see that they exhibit differences 
in their $T$-dependence parameters that are noticeable enough to suggest that using a single choice
of a rate for both of them would not be accurate enough. 
In any event, both polar molecules indicate strong increases in the elastic cross sections as the energy tends to zero for the existence of
enhanced cross section in that range of collision energies due to the presence of scattering from a dipole potential.

\begin{figure}
\begin{center}
\includegraphics[width=.45\textwidth]{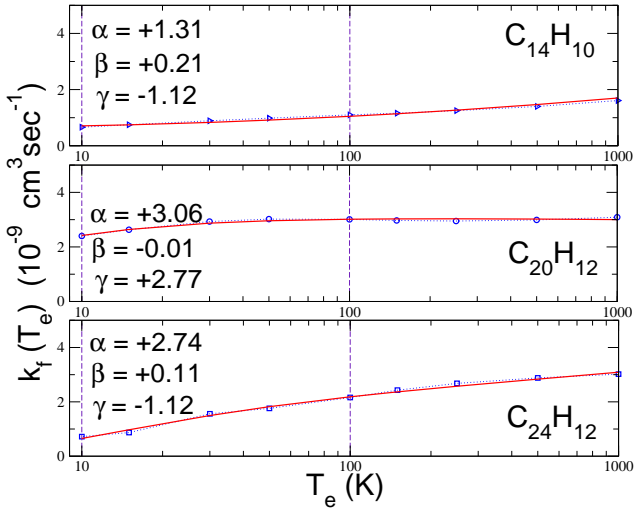}
\caption{Computed rates of electron attachment to three different polycondensed aromatic ring molecules. Top panel: anthracene;
middle panel: perylene; bottom panel: coronene. Note that $\alpha$ is in units of $10^{-9}$ cm$^3/$s. See main text for further details.\label{fig2}}
\end{center}
\end{figure}

The situation changes quite markedly when we now move to the examples of the polycyclic hydrocarbons
without permanent dipoles which are presented by the rates in Fig.\ref{fig2}.

The following comments could be made from the data reported in the figure.
\begin{itemize}
\item[(i)] All three examples show a different temperature dependence from those shown in Fig.\ref{fig1}, since we now
see that the rates are lowest at vanishing energies and increase with increasing $T$.
\item[(ii)] The sizes of all cross-sections are around $10^{-9}$ cm$^3$ s$^{-1}$, i.e. smaller than those for
polar molecules even if the number of C atoms is now larger.
\item[(iii)] The fitting parameters change along the series, but the $\alpha$ value, somewhat proportional to the rate size, does not increase with
increasing the number of C atoms. All the cross sections are fairly similar in size and behaviour along the small
series of PAHs shown in the figure.
\end{itemize}

One should furthermore remember that the present treatment of the attachment rates (see Eqns.(\ref{2Bkform}) to (\ref{kf_le_kTNI})) implies
that the TNI formations and the threshold resonances (virtual states) are the dominant gateways to the efficient energy dissipations paths that lead
to stable anionic formation. Thus, the attachment processes occur more efficiently at specific energy windows where such states are formed.
Additionally, since no VFR processes are directly included during the scattering calculations, we consider the contributions from vibrational Feshbach
resonances are possible (albeit conjecturally) at energies near threshold where our rates are already enhanced by the virtual state effects
(e.g. see \citealt{Carelli2012c}). 

The present calculations indicate that the chemical features of the involved PAHs do bear on the efficiency of the electron attachment 
rates obtained from the modelling reported in the present work.
The presence of a permanent dipole moment, in fact, is a significant rate enhancer, especially at the lower temperatures where non-polar
PAHs are found by our model to have lower rates. Furthermore, we see that all our computed rates, although considered to be upper bound to the
true rates for such systems, are consistently lower than the estimates from phase-space theory employed in earlier work 
for such quantities \citep{Wakelam2008}.

One possible reason for it could be that the phase-space considerations are not able to obtain the actual probabilities for the collisional formation of TNIs transition state and therefore may over-estimate formation efficiencies, as also surmised by \citet{Herbst2008}.

It is therefore interesting to see what would be the effect of the present results on the gas-phase models for the chemistry of dense interstellar
clouds, which can now employ attachment rates from ab initio calculations. This is the next aspect of the present work that we
discuss in the following section.  

\section{Testing our attachment rates with an evolutionary model}
To now assess the effects of the new rates for the electron attachment processes on the evolution of the corresponding abundances in the ISM environments we have implemented a pseudo time-dependent dense cloud model based on the work of \citet{Wakelam2008} 
 following their parameters and their recommendations for all the other rates involved in the fuller chemical network.
We discuss below the comparison between the results for the overall chemical evolutions using the single value of the rate coefficient that they assume for the PAH electron attachment process \citep{Wakelam2008} and the results obtained by employing the three rates proposed in this paper and individually obtained via the \emph{ab initio} calculations that we discussed in the previous sections.

For the rate coefficient $k_\mathrm{EA}$,
WH2008 suggest two values already found in the literature that are $k_ {fo}=10^{-7}(N_\mathrm{C})^{3/4}$ cm$^3/$s \citep{Omont1986}
and $k_{fa}=1.2\times10^{-7}(N_\mathrm{C})^{1/2}$ cm$^3/$s \citep{Allamandola1989}, where in both cases $N_\mathrm{C}$ is the number of carbon atoms that form the PAH. 
Their model assumes $N_\mathrm{C}=30$, and they arbitrarily decided to use the first formula as indicated in \citet{LePage2001}.

What we are therefore presenting below is a test aimed at determining the influence of using more accurate electron attachment rate coefficients on the  evolution of the ISM model. In particular, we track the evolution of the ISM (i) first with the aforementioned 
$k_\mathrm{fo}$ rate coefficient and then with the same set of other rates, but changing that of the PAHs  for (ii) anthracene $k_\mathrm{fA}$,  (iii) perylene $k_\mathrm{fP}$, and (iv) coronene $k_\mathrm{fC}$. 
All the other model parameters we adopt are the same as those of WH2008, namely $n_\mathrm{H2}=10^4$ cm$^{-3}$, 
a gas temperature of $T=10$ K, a cosmic-ray ionization rate of $\zeta_\mathrm{CR}=1.3\times10^{-17}$, a visual extinction
$A_\mathrm{v}=10$ s$^{-1}$, a PAH fractional abundance of $3.07\times10^{-7}$, and an average PAH size of 
$4\,\mathrm{\AA}$. 
We have selected  the initial species abundances given by the model EA2 (see Table 1 in WH2008), which is a high-metallicity  scenario
based on the observations of the dense cloud $\zeta$ Ophiuchi, except for He which is set to $n_\mathrm{He}=0.09$
compared to the total hydrogen abundance. More details on the full choice of parameters can be found in WH2008.

\begin{figure}
\begin{center}
\includegraphics[width=.35\textwidth, angle=-90]{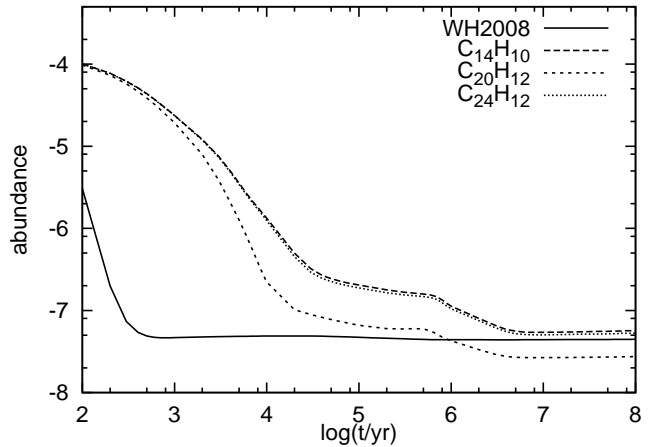}
\caption{The evolution of the total  abundances of the negatively charged species with time normalized to the total hydrogen number density
(i.e. $2\times10^4$ cm$^{-3}$). 
We compare the rates from the original WH2008 model that used their selection of $k_\mathrm{f}$ electron attachment rates (solid) 
with the cases: the anthracene rate (long-dashed), the perylene rate (dashed), and the
coronene rate (dotted). See text for further details. 
 \label{EA2_negative}}
\end{center}
\end{figure}

\begin{figure*}
\begin{center}
\includegraphics[width=.30\textwidth, angle=-90]{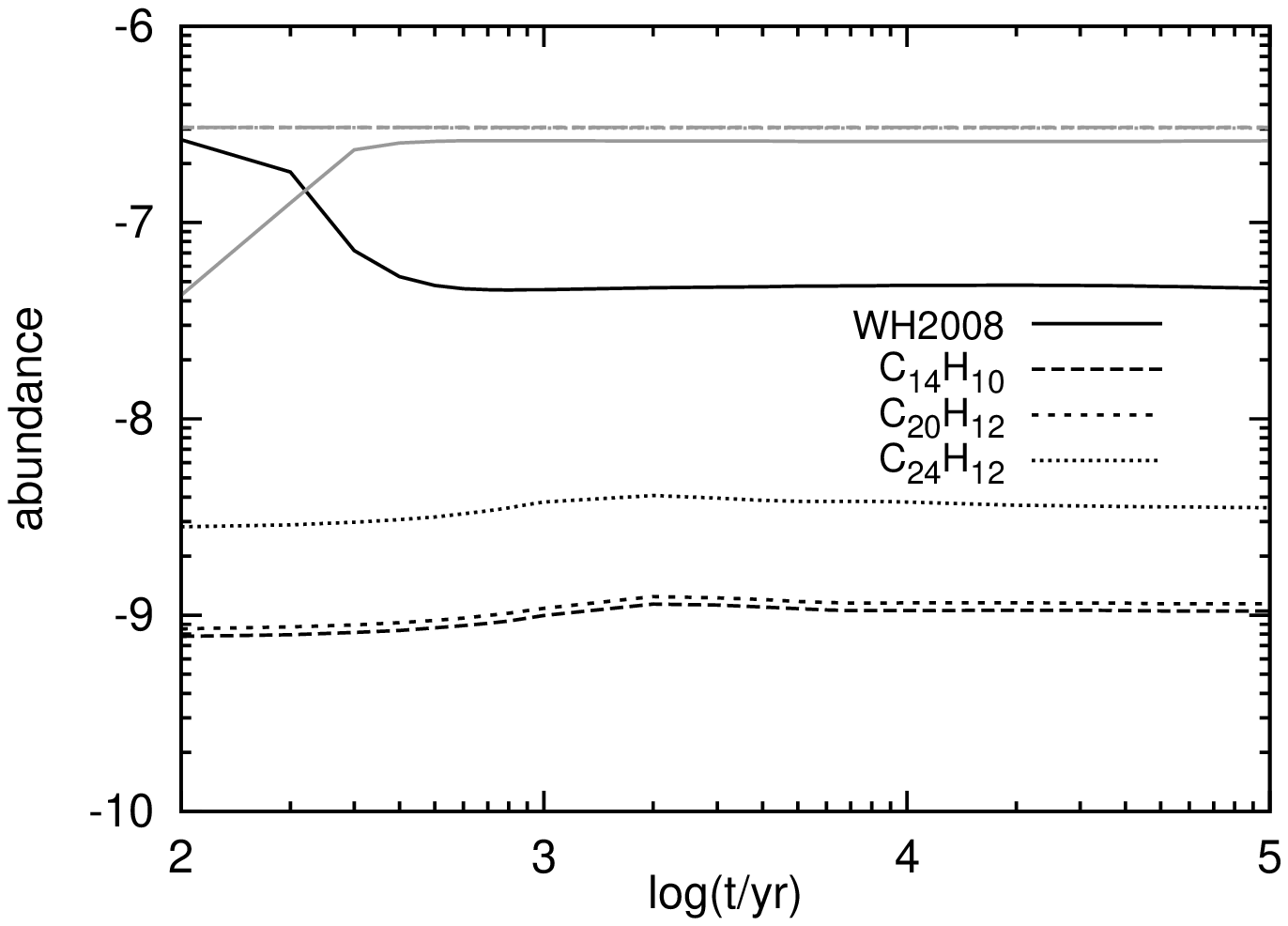}
\includegraphics[width=.30\textwidth, angle=-90]{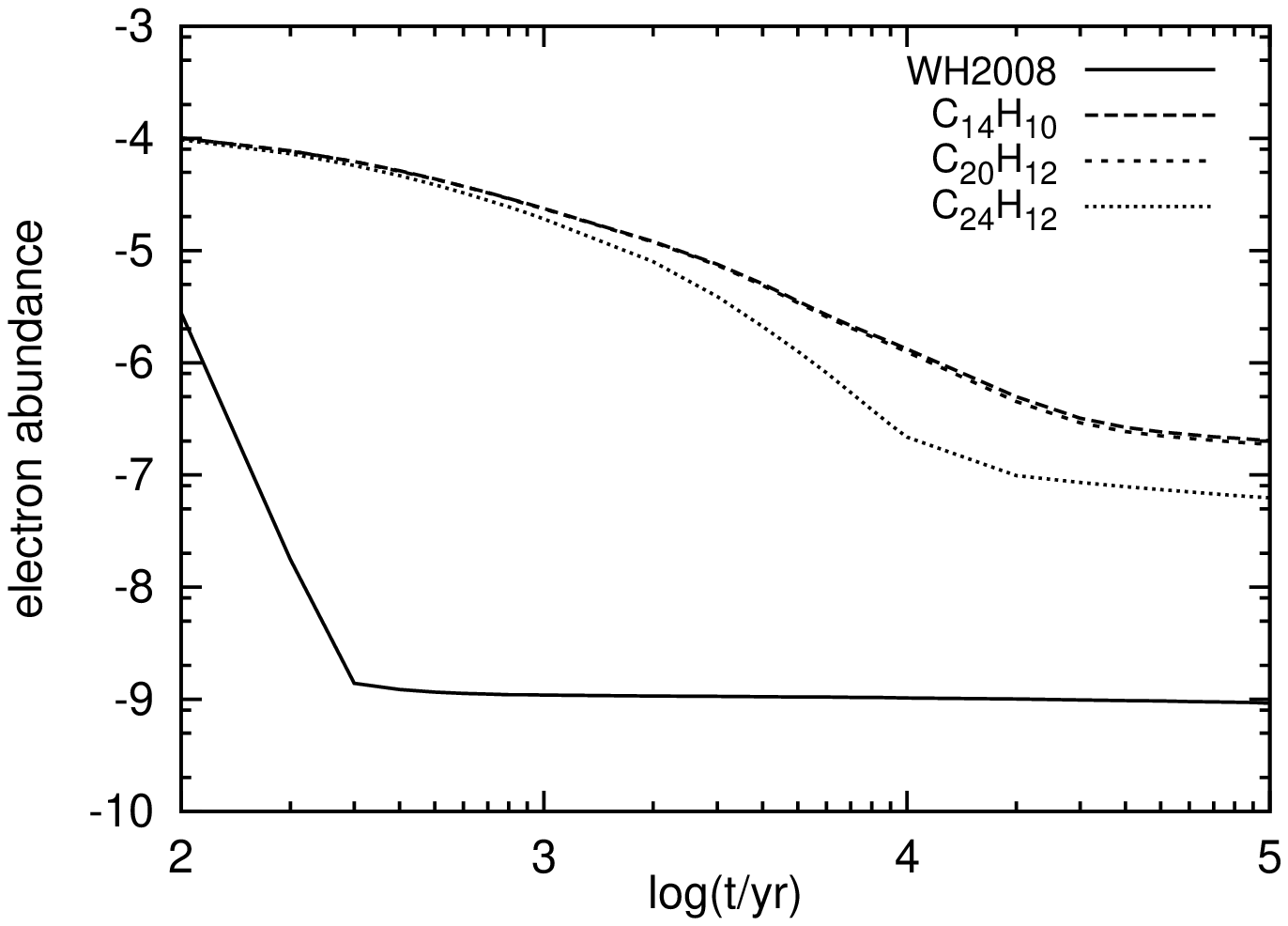}
\caption{Left panel: time evolution of the abundances of PAH (grey) and PAH$^-$ (black) normalized to the total hydrogen number density
(i.e. $2\times10^4$ cm$^{-3}$). 
We compare the WH2008 electron attachment rates (solid) with those obtained using instead. The anthracene rates (long-dashed), the perylene rates (dashed), and the coronene rate (dotted). The lines that represent the PAH densities overlap, except those from  WH2008. 
Right panel: electron number density time evolution as given by  the same set of models. The two panels
have different vertical scales. \label{EA2_PAH}}
\end{center}
\end{figure*}

The results of our simulations are displayed in Fig.\ref{EA2_negative} and Fig.\ref{EA2_PAH}. The first one
describes the evolution of the ionization fraction as the sum of the number densities of the negative species
at a given time. WH2008 had shown a fast decrease in the total negative charge until it reaches  steady state
after $10^3$ yr, while the present models have a slower decrease in time and  reach the steady-state on 
a much longer time scale, i.e. in $10^6$ yr or more. 
The effect of the new calculations is also shown in the figure and is rather clear: in the present model the PAHs manage to ``soak up''
 and capture fewer 
electrons when compared to the results from  \citet{Wakelam2008}. Their selected single value for the attachment rate coefficient is higher and thus indicates a more effective mechanism, and their modelling surmises the existence of a more efficient electron-attachment process than found here by \emph{ab initio} study. It then follows that the formed PAH anions rapidly recombine with the ions, thereby  obtaining a more neutral gas
as shown in Fig.\ref{EA2_negative}. On the other hand, when we replace the WH2008 rate coefficient with our separate rates for different 
molecules we find that the neutralization effect is less important, thus obtaining a less neutral gas once the steady state is reached.

In a similar fashion, the data in Fig.\ref{EA2_PAH} show the evolution of the PAH abundances over the time range $t=[10^2, 10^5]$ yr
and, even in this case, the effect of the new rate coefficients is evident. We obtain a larger separation between
the abundances of PAH and PAH$^-$ , a quantity that obviously increases when the attachment rate coefficients decrease with respect to the single value chosen in \citet{Wakelam2008} .

\section{Conclusions}
\label{conclusions}
In the present work we have endeavoured to analyse in some detail how one can realistically generate electron-attachment cross sections over a broad
range of energies for gas-phase polycondensed aromatic molecules, which is a specific subset of those molecules chosen to be representative
of the more general behaviour of PAHs in the interstellar and circumstellar environments. In particular, we selected a dynamical model that
employs the above cross sections to obtain attachment rates over a range of temperatures that are of direct interest in the ISM environments. The corresponding rates are essentially considered to be upper bounds to the true attachment rates, but are nevertheless found to already be less
than those employed in other evolutionary models, e.g. see \citet{Wakelam2008}, where no direct scattering calculations were used to generate
attachment rates.

Additionally, the parametric representation of computed rates is chosen to be the one usually employed in the literature to describe 2B processes in the ISM
\citep{Woodall2007} and is found by our present work to depend on at least two chemical properties of the aromatic molecules:
(i) the presence of a permanent dipole moment, a features that greatly increases the size of the attachment rates, especially at the
threshold temperatures; and (ii) a minor effect of the number of C atoms or of the number of condensed rings in the molecular system.

The capture rates were further employed to model the time evolution within a pseudo time-dependent dense cloud description that follows
the one given earlier by \citet{Wakelam2008}. In particular, we tried to analyse the effects of the attachment rate values produced by the 
present work on the evolutions of negatively charged species with time (e.g. see Fig.\ref{EA2_negative}) and the evolutions of PAH/PAH$^-$ 
abundances throughout the same model, as well as the electron number density changes over the same time period (see Fig.\ref{EA2_PAH}).

The corresponding calculations suggest rather clearly that:
\begin{itemize}
\item[(i)] the attachment rates are lower, at least over the relevant range of temperatures, than the values previously employed;
\item[(ii)] they depend on the chemical properties of the specific PAH being considered, especially when comparing anthracene and coronene;
\item[(iii)] the electron densities remain higher over a much broader range of time and reach a steady state much later that thought 
	before \citep{Wakelam2008};
\item[(iv)] the ``soaking-up'' power of the PAHs we have considered here turns out to be less than expected, hence reducing the efficiency
of the neutralization processes. 
\end{itemize}

Although we feel that further numerical experiments are needed in order to more extensively include a broader variety of polar PAHs within the
evolutionary model, the set of exemplary molecules presented here are already making it clear to us that chemical variety plays a role that
is bigger than 
previously expected and that a molecule-specific modelling of the role of electrons in evolutionary studies should be sought as much as possible. 

In conclusion we postulate here that the formation of a variety of TNIs and of zero-energy virtual states are all features included in our scattering calculations and are instrumental mechanisms in yielding  the final electron attachment rates that we in turn have employed for the evolutionary studies reported in the present work.

\begin{acknowledgements}
The computational support from the CASPUR Consortium is gratefully acknowledged, as well as the financial
support from the PRIN 2009 research network. One of us (T.G.) thanks the CINECA Consortium
for awarding a postdoctoral grant during which this work was carried out. We are also grateful to the referee for
convincing us to work on clearer presentation of our results.
\end{acknowledgements}

\bibliographystyle{aa}      
\bibliography{mybib} 
\end{document}